\documentclass[12pt]{article}
\usepackage{amsmath,amsfonts,amssymb,a4,epsfig,color}

\newcommand{\R}{{\mathbb{R}}}

\newcommand{\N}{{\mathbb{N}}}

\def\ha{\frac{1}{2}}

\def\ra{\rightarrow}
\def\preuve{\begin{proof}} 
\def\ga{\alpha}
\def\gb{\beta}

\def\gl{\lambda}

\def\OPD{pseudo-differential operator}

\newtheorem{lemm}{Lemma}

\newtheorem{rem}{Remark}

\newtheorem{theo}{Theorem}

\newenvironment{demo}{\noindent {\it Proof.--}
      \begin{quotation}\noindent}{\end{quotation}\hfill$\square $}

\begin{document}

\title{The action of  \OPD s on functions harmonic outside 
a smooth hyper-surface}
\author{Louis Boutet de Monvel\footnote{Institut Math\'ematique de
    Jussieu,
 Universit\'e Pierre et Marie Curie}~~~\& ~~
Yves Colin de Verdi\`ere\footnote{Institut Fourier,
 Unit{\'e} mixte
 de recherche CNRS-UJF 5582,
 BP 74, 38402-Saint Martin d'H\`eres Cedex (France);
yves.colin-de-verdiere@ujf-grenoble.fr}}


\maketitle

The goal of this note is to describe the action of \OPD s
 on the space ${\cal H}$  of $L^2$ functions 
which are harmonic  outside a smooth closed 
hyper-surface $Z$ of a compact  Riemannian
manifold without boundary  $(X,g)$ and whose traces from  both sides of $Z$
co\"incide. We will
 represent these $L^2$  harmonic functions as harmonic extensions
 of functions  in the Sobolev space  $H^{-1/2}(Z) $ by
 a Poisson operator ${\cal P}$.
The main result says that,  if $A$ is a \OPD ~of degree $d<3$,  the operator
$$
 B={\cal P}^\star  \circ A \circ {\cal P} ~
$$
is a \OPD~ on $Z$  of  degree $d-1 $  whose principal symbol of degree
 $d-1$  can be computed by integration of the principal symbol of $A$
 on
 the co-normal bundle of $Z$.

 These ``bilateral'' extensions are  simpler (at least
 for the Laplace operator) 
 than the ``unilateral'' ones whose study is the theory of
 \OPD s on manifolds with boundary (see \cite {Cal,LBM1,LBM2,Grub,Taylor}).

\section{Symbols}
The following classes of symbols are defined in the books
\cite{Grub},
 sec. 7.1,   and in \cite{Ho}, sec. 18.1. A {\it symbol of degree}
 $d$ on $U_x\times \R^n_\xi$ where $U$ is an open set in $\R^N$ is a
 smooth
 complex valued function $a(x,\xi)$ on $U\times \R^n$ which satisfies
 the following estimates:
for any multi-indices $(\ga,\gb)$, there exists a constant
 $C_{\ga,\gb}$ so that \[ | D^\ga _x D_\xi ^\gb a (x;\xi)|
 \leq C_{\ga,\gb} (1+\| \xi \|)^{d-|\gb|}~.\]
The symbol $a$ is called {\it classical} if $a $ admits
 an expansion $a \sim \sum_{l=0}^\infty  a_{d-l}$ where  $a_j$ 
 is homogeneous of degree $j$ ($j$ an integer) 
for  $\xi\in \R^n$ large enough; more precisely,
 for any $J\in \N$,  $ a-\sum_{j=0}^J a_{d-j} $  is a symbol of degree $d-J-1$.

We will need the 
\begin{lemm}\label{lemm:intsymb} 
If $a (x;\xi,\eta )$ is a  symbol of degree $d<-1$ defined
 on $U_x\times \left( \R^{n}_{\xi}\times \R_\eta \right)$,
  $ b(x;\xi )=\int _\R a(x;\xi,\eta ) d\eta$ is  a  symbol
 of degree $d+1$ defined on $U_x\times \R^{n}_\xi$. Moreover, 
 if $a$ is classical, $b$ is also classical and the homogeneous
 components of $b$ 
are given for $l\leq d+1$, by
 $ b_l(x;\xi )=\int _\R a_{l-1}(x;\xi,\eta ) d\eta$
\end{lemm}

\section{A general reduction Theorem for \OPD s}
We  choose local coordinates in some neighborhood of a point
 in $Z$ denoted $x=(z,y)\in \R^{d-1} \times \R$,  so that  $Z=\{ y=0
 \}$.
  We denote by $(\Omega _j ,~ j=1,\cdots N)$  a finite cover of $Z$ by
 such charts and denote by $\Omega _0$ an open set disjoint from $Z$
 so that $X=\cup_{j=0}^N \Omega _j $. We  choose the charts $\Omega
 _j$
 so that the densities $|dz|$ and $|dx| $ are the Lebesgue measures.

If $X$ is a smooth manifold, we denote by  ${\cal D'} (X)$ the space
 of generalized functions  on $X$ of which the space of
 smooth functions on $X$ 
is a dense subspace. We assume that $X$ and $Z$
 are equipped with smooth densities $|dx| $ and $|dz|$.
 This allows to identify generalized functions  with  Schwartz
 distributions, i.e. linear functionals 
  on test functions; this duality extending the $L^2$ product
 is denoted by  $\langle | \rangle $. 
We introduce the extension operator
 ${\cal E}:{\cal D'}  (Z) \ra {\cal D'}(X)$ sending the distribution
  $f $ to the distribution  $f\delta (y=0)$ defined 
$$
\langle f\delta (y=0)|\phi(z,y)\rangle = \langle f | \phi(z,0)
\rangle $$
and its adjoint, the trace ${\cal  T }:C^\infty (X)\ra C^\infty (Z)$
 defined by $\phi \ra \phi_{|Z}$.
Let  $A$ be  a \OPD ~ on $X$: let us call $A_j$ the restriction
 of $A$ to test functions compactly supported in $\Omega _j$.  We will
 work
 with one of the $A_j$'s given by the following ``quantization'' rule 
$$ 
A_ju(z,y)=\frac{1}{(2\pi)^{d}}\int_{ \R^{2d}}
  e^{i\left( \langle z-z'|\zeta\rangle  + (y-y')\eta \right)}
a_j(z,y;\zeta,\eta ) u(z',y') dz' dy' d\zeta d\eta ~.
$$
So we have formally, using the facts that the densities
 on $X$ and $Z$ are given by the Lebesgue measures in these local coordinates: 
$$
{\cal T}\circ A_j\circ{\cal E}v(z)= \frac{1}{(2\pi)^{d}}
\int_{\R^{2d-1}} e^{i \langle z-z'|\zeta\rangle    }
a_j(z,0;\zeta,\eta ) v(z') dz'  d\zeta d\eta ~,
$$
which we can rewrite
$${\cal T}\circ A_j\circ{\cal E}v(z)=\frac{1}{(2\pi)^{d-1}}\int_{\R^{2d}}  
 e^{i\langle z-z'|\zeta\rangle  } b_j(z;\zeta ) v(z') dz'  d\zeta ~,
 $$
with
\begin{equation} \label{equ:symbol}
 b_j(z;\zeta ) =\frac{1}{2\pi}\int_\R  a_j(z,0;\zeta,\eta ) d\eta ~.
\end{equation}

We have the
\begin{theo}\label{theo:main}
 If $A$ is a \OPD~ on $X$ of degree $m<-1$ whose full symbol in the
 chart $\Omega _j$ is $a_j$, 
then  the operator ${\cal T}\circ A\circ{\cal E}$ is a \OPD~ on $Z$
  of degree $m+1$ whose  symbol is given in the charts $\Omega _j\cap
  Z$
 by Equation (\ref{equ:symbol}).
 \end{theo} 
 This is proved by looking at the actions on test functions compactly
 supported in the chart $\Omega _j,~j\geq 1$:  then we use Lemma \ref{lemm:intsymb}. 
 \begin{rem}
The principal symbol can be described in a more intrinsic way:
 let $z\in Z$ be given, from the  smooth densities on $T_zX$ and
 on $T_zZ$ given by $|dx|$ and $|dz|$, we get, using the Liouville
  densities, densities on the dual bundles  $T_z^\star Z$ and
  $T_z^\star X$.
 Let us denote by $\Omega^1(E)$ the
1-dimensional space of densities on the vector space $E$. From the
exact
 sequence
$$
0 \ra N^\star _z Z\ra T^\star _zX \ra T^\star _z Z \ra 0~,
$$
we deduce
$$ 
\Omega ^1 (T^\star X) \equiv \Omega ^1 (N^\star Z)\otimes 
 \Omega ^1 (T^\star Z )~
 $$
and a canonical density $dm (z)$  in  $\Omega ^1 (N_z^\star Z)$.
 The principal symbol of $B={\cal T}\circ A\circ{\cal E}$ is given
 in coordinates  by $b(z,\zeta)=(1/2\pi)\int_{N_z^\star Z} a(z;\zeta,\eta )dm(\eta) $.
\end{rem}

\section{ The ``bilateral''  Dirichlet-to-Neumann operator}
 We will assume that the local coordinates $x=(z,y)$  along $Z$ are
 chosen
 so that $g(z,0)=  h(dz) +dy^2$ and the Riemannian volume along $Z$ is
 $|dx|_g=|dz|_h |dy|$. We will choose the associated densities on $X$
 and $Z$.
 We will denote by $\Delta _g$ the Laplace-Beltrami operator on
 $(X,g)$
 as defined by Riemannian geometers (i.e. with a minus sign in front
 of
 the second order derivatives).

 If $f $ is given on $Z$, let us denote by ${\cal DN}(f)$
  minus  the sum of the interior  normal derivatives on both
 sides of $Z$ of  the harmonic extension $F$  of $f$;
this always  makes sense, even if the normal bundle of $Z$ is not orientable.
 We have the
\begin{lemm} \label{lemm:harm-ext}
 The distributional Laplacian of the harmonic extension  $F$
 of a smooth function $f$ on $Z$  is 
$\Delta_g F={\cal E}({\cal DN}(f))$.
\end{lemm}
\begin{demo} The proof is a simple application of the Green's formula:
 by definition of the action of the Laplacian on distributions, if
 $\phi $ is a test function on $X$, $ \langle \Delta_g F|\phi \rangle
 :=
 \langle  F |\Delta_g \phi \rangle $.
We can compute the righthandside  integral as an integral on
$X\setminus Z$
 using  Green's formula.
$$
 \int _{X\setminus Z}(F\Delta_g \phi -\phi \Delta _g F)|dx|_g=
 \int _Z (F\delta \phi - \phi \delta F )|dz|_h
 $$
where $\delta $ is   the sum of the interior normal derivatives from
 both sides of $Z$. Using the fact that $\Delta _g F =0 $ in
 $X\setminus Z$
 and $\delta \phi=0$, we get the result.
\end{demo} 

Denoting by $\Delta_g^{-1}$ the ``quasi-inverse''  of $\Delta_g$
defined
 by $\Delta_g^{-1}\phi_j =\gl_j^{-1} \phi_j$ for the eigenfunctions
 $\phi_j$ of $\Delta _g$  with non-zero eigenvalue $\gl_j$ and 
 $\Delta_g^{-1}1=0$, we have $f=\left({\cal T}\circ\Delta_g
   ^{-1}\circ{\cal E}\right) \circ {\cal DN}(f)~({\rm mod~
   constants})$.
 By Theorem \ref{theo:main}, the operator $B={\cal T}\circ\Delta_g
 ^{-1}
\circ{\cal E}$ is an elliptic self-adjoint \OPD~ on $Z$.
 The operator ${\cal DN}$ is a right inverse of $B$  modulo smoothing
 operators and hence also a left  inverse modulo smoothing operators.
 So that ${\cal DN}=B^{-1}$ is an elliptic self-adjoint of principal
 symbol the inverse 
$$
 \frac{1}{2\pi}\int_\R  (\|\zeta\|_h^2+\eta^2)^{-1} d\eta = 
\frac{1}{2\| \zeta \|_h} ~,
$$
namely $2\| \zeta \|_h$. Hence
\begin{theo}\label{theo:dn} 
 The bilateral Dirichlet-to-Neumann${\cal DN}$ is   a self-adjoint
  elliptic  \OPD~  of  degree $1$ on 
 $L^2(Z,|dz|)$ and of principal symbol $2\| \zeta \|_h$.
 The kernel of  ${\cal DN}$ is the space of constant functions.  
\end{theo}
The full symbol of  ${\cal DN}$ can be computed in
 a similar way from the full symbol of the resolvent $\Delta_g^{-1}$ along $Z$.

\section{The Poisson operator}

Let $A$ be an \OPD~ on $X$ of principal symbol $a$. We are
 interested to the restriction to the space ${\cal H}$ of 
 the quadratic form $Q_A(F)=\langle AF |F \rangle $ associated to $A
 $.
 We will parametrize ${\cal H}$ as harmonic extensions  of functions
 which are in $H^{-\ha}(Z)$ by the so-called Poisson operator denoted
 by ${\cal P}$; the pull-back $R_A$  of $Q_A$ on $L^2(Z)$ is defined by
$$
 R_A (f)=\langle A{\cal P}f |{\cal P}f \rangle =
 \langle {\cal P}^\star A{\cal P}f |f \rangle~.
$$
The goal of this section is to compute the operator
 $B={\cal P}^\star A  {\cal P}$ associated to the quadratic form $R_A$.

From Lemma \ref{lemm:harm-ext}, we have,   modulo smoothing operators,
 $$
  {\cal P}= \Delta_g^{-1}\circ {\cal E}\circ {\cal DN }~.
  $$
Hence
$$ 
B= {\cal DN }\circ \left[ {\cal T}\circ \left(\Delta_g^{-1}\circ A 
\circ\Delta_g^{-1}\right)\circ {\cal E}\right] \circ {\cal DN }~.
$$
   
The operator  $ \Delta_g^{-1}\circ A \circ\Delta_g^{-1}$
 is a \OPD~ of principal symbol $a/(\| \zeta\|_h^2 +\eta^2)^2$ near
 $Z$.

  Applying Theorem \ref{theo:main} to the inner bracket 
 and Theorem \ref{theo:dn}, we get the:
\begin{theo} \label{theo:poisson}
If $A$ is a \OPD ~ of degree $d<3$ on $X$ and ${\cal P}$
the Poisson operator associated to $Z$, the operator
 $B={\cal P}^\star A  {\cal P}$ is a \OPD ~ of degree $d-1$ on $Z$
of principal symbol
$$
 b(z,\zeta)=\frac{2}{\pi} \| \zeta \|_h^2 \int _\R
 \frac{a(z,0;\zeta, \eta)}{(\| \zeta \|_h^2+ \eta ^2)^2}d\eta ~.
$$
\end{theo}
\begin{rem} Note that if $A$ is a pseudo-differential
 operator without the transmission property, 
the operator $A\circ {\cal P}$ may be ill-behaved and have disagreeable
 singularities along $Z$; however ${\cal P}^*A{\cal P}$ is always a
 good pseudo-differential operator on $Z$.
\end{rem}

\bibliographystyle{plain}

\end{document}